\documentclass[prx,reprint,noshowpacs,superscriptaddress,floatfix,letterpaper]{revtex4-2}
\usepackage{amsmath,amssymb,amsbsy,amsfonts,amsthm,mathrsfs,amssymb}
\usepackage{bm,bbm,graphicx}
\usepackage{float}
\usepackage{epsfig}
\usepackage{color}
\usepackage[dvipsnames]{xcolor}
\usepackage[colorlinks=true,citecolor=Cerulean,linkcolor=RubineRed,urlcolor=Cerulean]{hyperref}
\usepackage{silence}
\usepackage{soul}
\usepackage[normalem]{ulem}
\usepackage{xfrac}
\usepackage[version=4]{mhchem}
\usepackage{stackrel}
\usepackage{sidecap}
\sidecaptionvpos{figure}{c}

\DeclareFontFamily{OT1}{pzc}{}
\DeclareFontShape{OT1}{pzc}{m}{it}{<-> s * [0.900] pzcmi7t}{}
\DeclareMathAlphabet{\mathpzc}{OT1}{pzc}{m}{it}

\newcommand{\sm}[1]{\textcolor{blue}{#1}}
\newcommand{\hyd}{{\mathrm{hyd}}}
\newcommand{\tot}{{\mathrm{tot}}}
\newcommand{\conf}{{\mathrm{conf}}}
\begin{document}

\title{Pareto optimal fronts of kinetic proofreading}
\author{Davide Chiuchiu}
\thanks{Equal contributions.}
\affiliation{Biological Complexity Unit,\ Okinawa Institute of Science and Technology Graduate University,\ Onna, Okinawa 904-0495, Japan}

\author{Shrabani~Mondal}
\thanks{Equal contributions.}
\affiliation{Biological Complexity Unit,\ Okinawa Institute of Science and Technology Graduate University,\ Onna, Okinawa 904-0495, Japan}
\affiliation{ Department of Chemistry, Physical Chemistry Section, Jadavpur University, Kolkata, 700032, India}

\author{Simone Pigolotti}
\email[]{simone.pigolotti@oist.jp}
\affiliation{Biological Complexity Unit,\ Okinawa Institute of Science and Technology Graduate University,\ Onna, Okinawa 904-0495, Japan}

\begin{abstract}
Biological processes such as DNA replication, RNA transcription, and protein translation operate with remarkable speed and accuracy in selecting the right substrate from pools of chemically identical molecules. This result is obtained by non-equilibrium reactions that dissipate chemical energy.  It is widely recognized that there must be a trade-off between speed, error, and dissipation characterizing these systems.  In this paper, we quantify the trade-off between speed, error, and dissipation using tools from mathematical optimization theory. We characterize the Pareto optimal front for a generalization of Hopfield's kinetic proofreading model, which is a paradigmatic example of biological error correction. We find that models with more proofreading steps are characterized by better trade-offs. Furthermore, we numerically study scaling relations between speed, accuracy, and dissipation on the Pareto front.

\bigskip

\hfill%
\begin{minipage}{12cm}
{\footnotesize Subject Areas: Biological Physics, Statistical Physics}
\end{minipage}

\end{abstract}

\maketitle

\section{Introduction}
In living cells, information encoded in the DNA is constantly transcribed into RNA, which in turn is translated into proteins. Moreover, this information must be reliably copied into new DNA before a cell division occurs. These processes are fundamental in biology; their speed and accuracy have a profound impact on the organism fitness \cite{ZAHER2009,Johnson1993, SAVIR2013471}. Evolution must therefore have shaped these processes to achieve high performance.

It has long been recognized that the accuracy of these reactions, let alone their speed, can not be achieved close to thermodynamic equilibrium \cite{pauling1957}. This means that information-replicating enzymes must necessarily catalyze non-equilibrium chemical reactions. A paradigmatic example is the kinetic proofreading scheme independently proposed by Hopfield \cite{Hopfield4135} and Ninio \cite{NINIO1975587}. Kinetic proofreading can lead to an accuracy significantly higher than the equilibrium one, at the cost of dissipating chemical energy. This idea profoundly impacted our understanding of error correction in biology. 

The concept of kinetic proofreading sparked an interest in experimentally characterizing the reaction networks of replicating enzymes and measuring their kinetic rates \cite{doi:10.1098/rstb.2016.0182, doi:10.1021/bi100556m, doi:10.3109/10409239309086792}. These studies have revealed that these networks usually include several intermediate steps and are more complex than the kinetic proofreading model. Unfortunately, these intermediate states are usually very short-lived \cite{doi:10.1098/rstb.2016.0182} and, therefore, hard to observe experimentally.

Alternative approaches attempt to characterize the performance of replicating enzymes without relying on knowledge of the underlying kinetic details. Often, these approaches invoke some form of optimality principle \cite{savir2013ribosome}. For example, speed and accuracy of {\em in vitro} translation are simultaneously affected by altering the concentration of magnesium ions in the assay \cite{magnus_pnas, vorstenbosch1996g222d}. This observation has led to the idea that a trade-off exists between accuracy and speed. Besides speed and accuracy, energy dissipation is another important property that biological systems can optimize \cite{Bennett1979}. Several studies have focused on trade-offs among speed, accuracy, and dissipation in biological error correction \cite{Savageau1979, EHRENBERG1980333, Murugan12034, Wong2018, Qyu2020, hartich2015}. However, these trade-offs might depend on the choice of the parameter being tuned \cite{Rao_2015,Pigolotti2016}. More fundamental bounds on speed, error, and dissipation of replicating enzymes are set by the second law of thermodynamics \cite{PhysRevX.5.041039,SEIFERT2018176, PhysRevLett.123.038101}. But despite their theoretical interest, these bounds are usually far from the operating regimes of replicating enzymes. This suggests that the formulation of optimality principles and trade-offs requires concrete implementations of the reaction networks. Several studies investigated tradeoffs between pairs of observables \cite{banerjee2017elucidating, mallory2019trade}. In particular, a recent study \cite{doi:10.1098/rsif.2021.0883} has theoretically derived a trade-off between error and dissipation. However, the modeling assumptions made in \cite{doi:10.1098/rsif.2021.0883} are such that the process speed can be varied independently of error and dissipation and is therefore not subject to a tradeoff.

In this paper, we numerically study trade-offs between speed, error, and dissipation using the concept of a Pareto front. We consider, as a paradigmatic example, a multi-step generalization of the original Hopfield model. We find that the model performance significantly improves at increasing the number of intermediate proofreading steps. We also characterize scaling relations between speed, error, and dissipation on the Pareto front.
 
\section{Pareto optimal front}\label{sec:pareto}

\begin{figure}
	\centering
	\includegraphics[width = \linewidth]{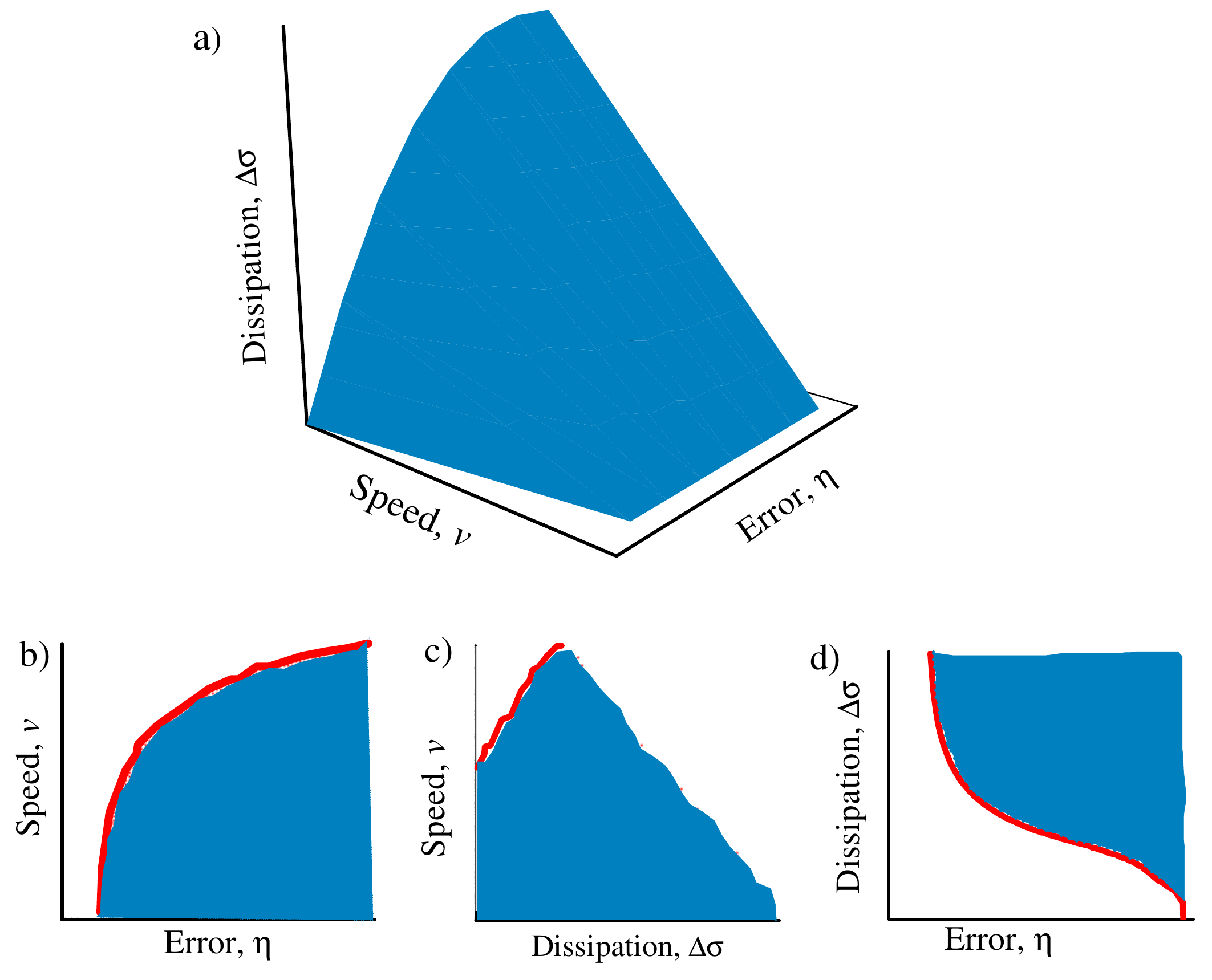}
	\caption{Feasible values of speed, error, and dissipation for the generalized Hopfield network with four proofreading steps. (a) Feasible configuration in the speed-error-dissipation phase space.
	Two-dimensional projections of the speed-error-dissipation phase space for (b) speed-error, (c) speed-dissipation, and (d) dissipation-error. 
	The Pareto optimal front is represented by a red curve. Model details and parameter values are discussed in Sections \ref{generalized_Hopfield_model} and \ref{sec5}.
	}
	\label{fig:pareto_fronts_an_illustative_example}
\end{figure}

We illustrate the idea of a Pareto front with an example in which we simultaneously optimize the speed $v$, the error rate $\eta$, and the dissipation per incorporated monomer $\Delta \sigma$, as a function of the kinetic rates. Kinetic constraints determine feasible combinations of speed, error, and dissipation. An example of such constraints is the fact that the reaction network is driven out of thermodynamic equilibrium by a finite energy budget. The feasible combinations $(v,\eta,\Delta \sigma)$ form a set, that we call $F$.   As an illustration, the set of feasible combinations for a generalization of the Hopfield model is shown in Figure \ref{fig:pareto_fronts_an_illustative_example}(a). The Pareto front $P$ is the subset of $F$ made up of the optimal configurations,  i.e., those in which one observable improves only at the expense of other observables \cite{miettinen2012nonlinear}. Here, by ``improvement of observable'' we mean an increase of the speed and a decrease of the dissipation and the error. Formally, a triplet $(v,\eta,\Delta \sigma)$ belongs to the Pareto front if there is no other triplet $(v',\eta',\Delta \sigma')\in F$ such that $v'\ge v$, $\eta'\le \eta$, and $\Delta \sigma'\le \Delta \sigma$. Feasible combinations of observables values that do not lie on the Pareto optimal front constitute sub-optimal solutions where a suitable parameter change can improve some observables without penalizing others.

For simplicity of illustration, we show feasible configurations and Pareto fronts for pairs of observables, see Figures~\ref{fig:pareto_fronts_an_illustative_example}(b)--(d). The definitions of feasible configurations and Pareto fronts for pairs of observables are obtained by marginalization from the three-dimensional case. For example, a pair $(v,\eta)$ is a feasible speed-error pair if $(v,\eta,\Delta\sigma)\in F$ for some $\Delta \sigma$. The corresponding Pareto front is the set of feasible configurations $(v,\eta)$ such that there are no other pairs $(v',\eta')$ such that $v'\ge v$ and $\eta'\le \eta$. Pareto fronts for pairs of observables are represented as red curves in Figures~\ref{fig:pareto_fronts_an_illustative_example}(b)--(d).

The Pareto front constitutes only a part of the boundary of the set of feasible configurations. In the example of Figure~\ref{fig:pareto_fronts_an_illustative_example}(c), the boundary on the right side of the figure is not part of the Pareto front since dissipation is not optimal along this line.

\section{Generalized Hopfield model}
\label{generalized_Hopfield_model}

\subsection*{Definition of the model}

\begin{figure*}
	\centering
		\includegraphics[width = \linewidth]{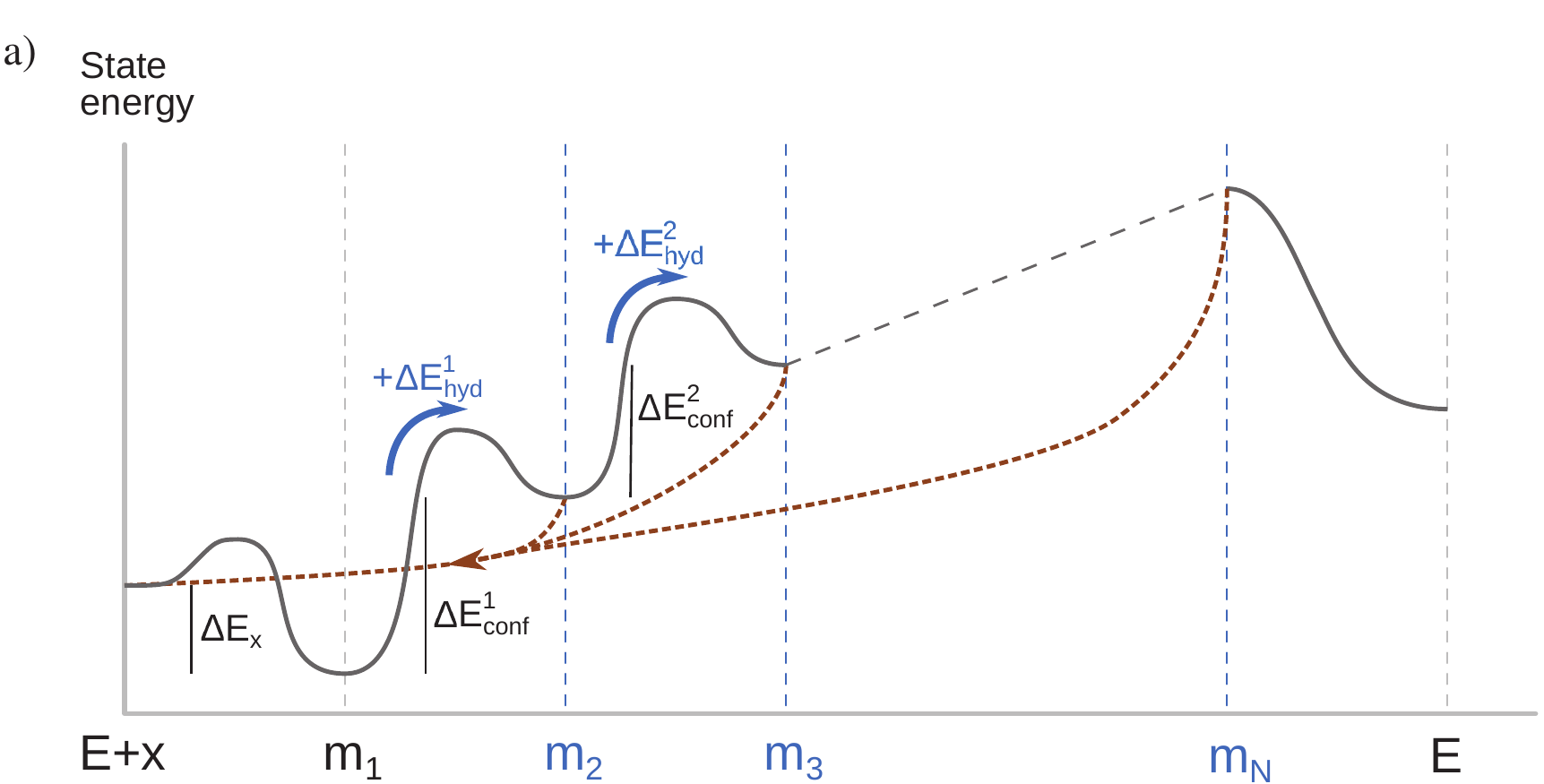}
		\includegraphics[width =\linewidth]{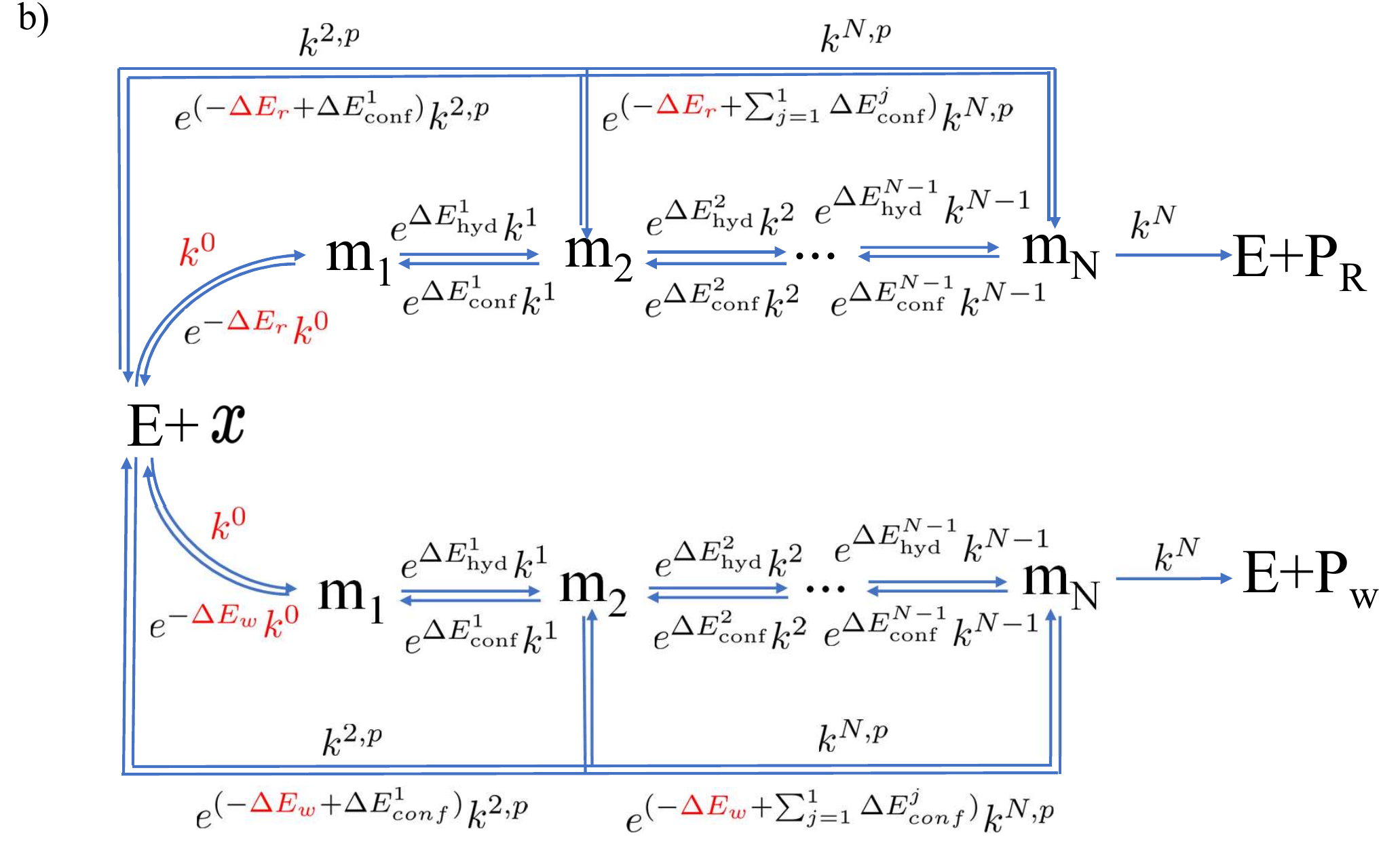}
	\caption{Generalized Hopfield model with $N$ intermediate steps. a) Energy landscape of the generalized Hopfield model. The depth $\Delta E_x$ of the first valley represents the binding energy of monomer $x$. The following states are characterized by positive energy increments $\Delta E_{\conf}^i$. Forward reactions are driven out of equilibrium by energies $\Delta E_{\hyd}^i$ originating from hydrolysis. We assume that neither $\Delta E_{\conf}^i$ nor $\Delta E_{\hyd}^i$ depend on the monomer type. The energy of the final state is not shown because we assume that the last reaction is fully irreversible. b) Chemical reaction network of the generalized Hopfield model. The enzyme goes through $N$ intermediate steps before incorporating monomer $x$. The reactions leading to the intermediate states $m_2\ldots m_N$ are assisted by an external nonequilibrium driving. In addition, $N-1$ proofreading pathways permit to reject the candidate monomers from the intermediate states $m_2\ldots m_N$. Rate constants and energy values that are kept fixed, rather than varied in the optimization process, are shown in red. }
	\label{fig:chemical_reactions_hopfield}
\end{figure*}

We study a generalized version of the Hopfield model \cite{Hopfield4135} including multiple proofreading steps. A replicative enzyme $E$ can bind to free monomers of type $x\in \{r,w\}$, where $r$ represents a right monomer and $w$ a wrong one. The enzyme may then reject the monomer based on its binding free energy $\Delta E_x$. In the incorporation pathway, the enzyme sequentially adopts $N$ configurations $m_i^x$, $i=1\dots N$ before the final incorporation of monomer $x$, where $i=1$ denotes the initial binding, see Figure~\ref{fig:chemical_reactions_hopfield}(a). 

We assume that the free energy changes $\Delta E^i_{\conf}$ when going from configuration $i$ to configuration $i+1$ do not depend on the monomer type. Here and in the following, all energies are expressed in units of the thermal energy $k_{B}T$, where $k_{B}$ is the Boltzmann constant and $T$ is the temperature of the environment. From the configurations  $i=2\dots N$, monomer $x$ can be discarded via a proofreading reaction that brings the replicative enzyme back to its initial free state. Proofreading rates depend on the energy of intermediate states and the monomer type. Similar to the original Hopfield model, the enzyme consumes chemical energy from hydrolysis to assist the  configuration changes. The enzyme absorbs this energy from a single hydrolysis event. However, we assume that this energy can be partitioned among the multiple steps leading to each of the intermediate states. Thus, the energy $\Delta E_{\hyd}^i$ consumed in going from the $(i-1)$th to the the $i$th configuration satisfies
\begin{equation}\label{eq:fixed_energy}
	\sum_{i = 1}^{N-1} \Delta E_{\hyd}^i = \Delta E_{\hyd}^{\tot}
\end{equation}
where $\Delta E_{\hyd}^{\tot}$ is the (fixed) total energy budget that the enzyme gains from one hydrolysis event.  We also assume that all reactions except the final incorporation are reversible. The reaction rates are determined by the energy landscape in Figure~\ref{fig:chemical_reactions_hopfield}(a). 

Specifically, the concentrations $[m_i^x]$ of the intermediate states evolve as
\begin{eqnarray}
	\begin{aligned}
			\frac{d[m_1^x]}{dt} = & k^{0} + e^{\Delta E_{\conf}^1} k^{1} [m_2^x]\\  
			&\, - \left(e^{\Delta E_{hyd}^1}\, k^{1} + e^{-\Delta E_x}\, k^0\right) [m_1^x]\\ 
				\frac{d[m_i^x]}{dt} = &\, k^{i,p} + e^{\Delta E_{\hyd}^{i-1}}\, k^{i-1} [m_{i-1}^x] + e^{\Delta E_{\conf}^i}k^{i} [m_{i+1}^x] \\ 	
			&\, -\left(e^{\Delta E_{\hyd}^i}\, k^{i} +  \, e^{[-\Delta E_x+\sum_{j=1}^{i-1} \Delta E_{\conf}^j]} k^{i,p}\right)[m_i^x] \\ 
			&+\left( e^{\Delta E_{\conf}^{i-1}}  k^{i-1} \right)[m_i^x]\\  
			& \mbox{for } i = 2\ldots N-1\\ 
				\frac{d[m_N^x]}{dt} = &\, k^{N-1,p} + e^{\Delta E_{\hyd}^{N-1}}\, k^{N-1} [m_{N-1}^x] \\ 
			&\, -\left(k^{N} +  e^{[-\Delta E_x+\sum_{j=1}^{N-1} \Delta E_{\conf}^j]}            k^{N-1,p} \right)[m_N^x]\\ 
			& + \left( e^{\Delta E_{\conf}^{N-1}} k^{N-1} \right)[m_N^x]  \, ,
	\end{aligned}\label{eq:hopfield_chemical_equations}
\end{eqnarray}
see Figure~\ref{fig:chemical_reactions_hopfield}(b). The binding energy $\Delta E_x$ depends on the monomer type, leading to different rejection rates for different monomer types. For simplicity, we assume that the concentrations of enzyme $[E]$ and monomers $[x]$ are maintained constant, and we implicitly incorporate them in the reaction rates involving free enzymes or monomers.

We compute the steady values of the concentrations $[m_i^x]$ from Eq.~\eqref{eq:hopfield_chemical_equations}. We then express the error $\eta$, the polymerization speed $v$, and the  dissipation per incorporated monomer $\Delta \sigma$ as
\begin{equation}
	\begin{aligned}
		\eta = &\, \frac{[m_N^w]}{[m_N^w] + [m_N^r]}\, ,\\
		v  = &\, k^N\left([m_N^r] + [m_N^w] \right)\, ,\\
		\Delta \sigma =&\, \frac{\sigma^r+\sigma^w}{v}\, ,
	\end{aligned}\label{eq:observables}
\end{equation}
where 
\begin{equation}
	\begin{aligned}
		\sigma^x = & k^0\left(1  - e^{-\Delta E_{x}}\left[m_1^x\right] \right) \log\left(\frac{1 }{e^{-\Delta E_{x}} \left[m_1^x\right]} \right) \\
		& + \sum_{i = 1}^{N-1} k^i
		\left(e^{\Delta E_{\hyd}^i} \left[m_i^x\right] - e^{\Delta E_{\conf}^i} \left[m_{i-1}^x\right] \right) \\
	&\times \log\left[\frac{e^{\Delta E_{\hyd}^i}  \left[m_i^x\right]}{ e^{\Delta E_{\conf}^i}   \left[m_{i-1}^x\right]} \right] \\
		& + \sum_{i = 1}^{N-1} 
        	k^{i, p} 	\left(1 -  e^{[-\Delta E_x+\sum_{j=1}^{i}\Delta E_{\conf}^j]}\left[m_i^x\right] \right) \\
	&\times 	\log\left[\frac{1}{e^{[-\Delta E_x+\sum_{j=1}^{i}\Delta E_{\conf}^j]}  \left[m_i^x\right]} \right] 
	\end{aligned}
\end{equation}
is the average entropy production rate depending on the monomer type. This definition of the average entropy production rate does not include the contribution of the irreversible final incorporation step,  as discussed in Ref. \cite{Rao_2015}. 

We take as free parameters the rate constants $k_1,k_2,\dots,k_N$. We assume that the initial binding rate $k_0$ is determined by diffusion and monomer concentration and thus can not be optimized. We therefore set $k_0=1$. Further free parameters are the hydrolysis energies at each step $\Delta E_{\hyd}^i$, $i=1\dots N-1$, and the intrinsic proofreading rates $k^{2,p}\dots k^{N,p}$. The total number of free parameters is thus $3N-2$, which reduces to $3N-3$ by considering that the hydrolysis energies are constrained by Eq.~\eqref{eq:fixed_energy}. 

In the optimization, we also impose that the direction of the average flux associated with each reaction is consistent with a process of polymer synthesis:
\begin{equation}
	\begin{aligned}
		1 -  e^{-\Delta E_x} [m_{1}^x] \geq & 0\\
		e^{\Delta E_{\hyd}^i} k^{i} [m_i^x] - e^{\Delta E_{\conf}^i} k^i [m_{i + 1}^x] \geq & 0 \\ 
		\mbox{for } i = 1\ldots N-1 \\
		k^{i, p} - e^{[-\Delta E_x+\sum_{j=1}^{i}\Delta E_{\conf}^j]} k^{i, p} [m_{i + 1}^x] \leq & 0 \\
		\mbox{for } i = 1\ldots N-1\, .
	\end{aligned}\label{eq:nonlinear_constraints}
\end{equation}
Additionally, we require that the enzyme has a fixed energy budget for each incorporated monomer via Eq.~\eqref{eq:fixed_energy}, and we impose that
\begin{equation}
	\frac{k^{i+1,p}}{k^{i,p}} < 1 \quad \mbox{for } i = 1\ldots N-1 \label{eq:linear_constraints}
\end{equation}
so that the states $m_i^x$ have increasing energy as shown in Figure \ref{fig:chemical_reactions_hopfield}. 

\begin{figure*}
	\includegraphics[width = \linewidth]{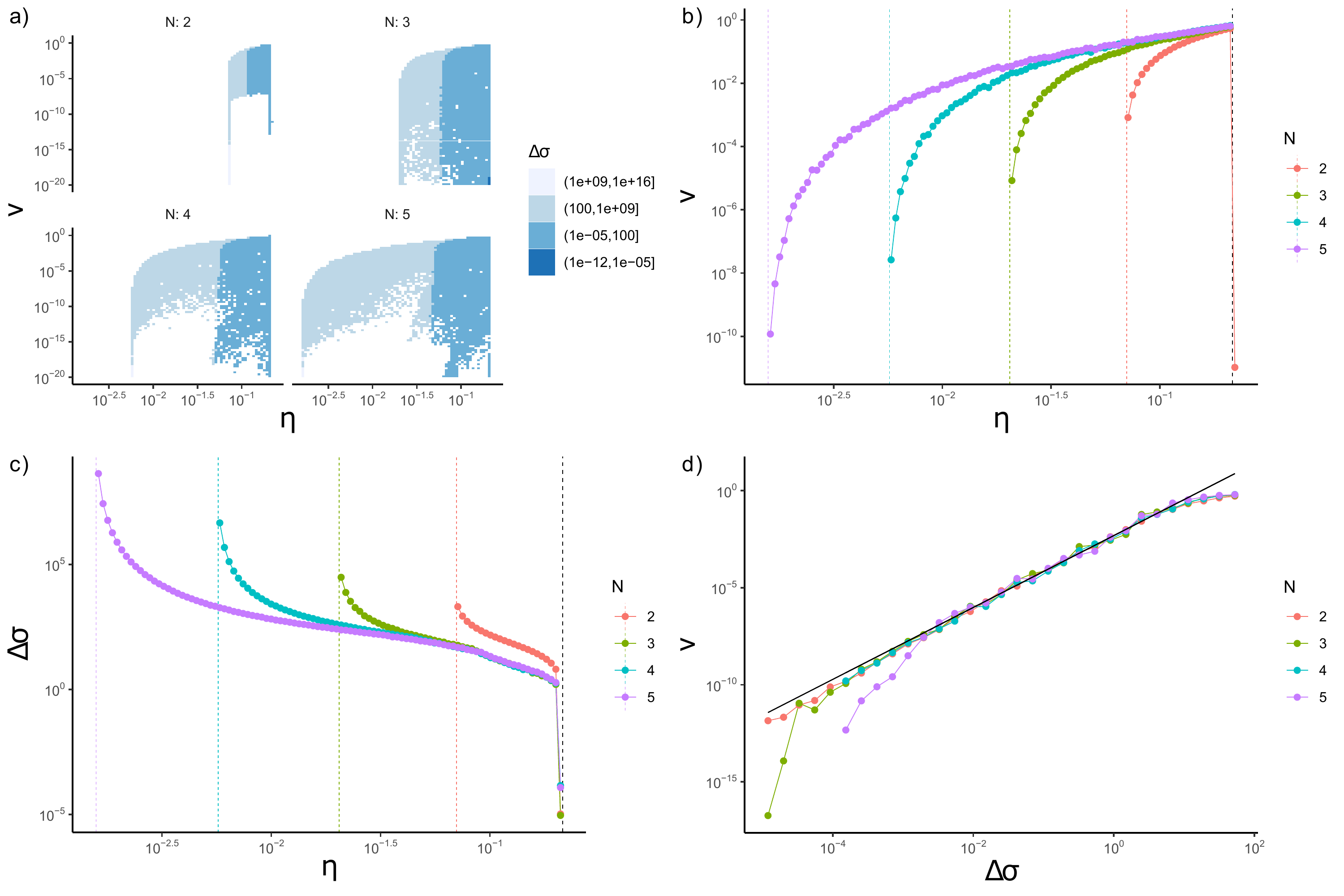}
	\caption{Additional proofreading steps improve Pareto optimal fronts.
		\sm{a}) Three-dimensional Pareto front between $\eta$, $v$ and $\Delta \sigma$ for different numbers of steps $N$.  Each blue point corresponds to the speed and error of particular sets of rate constants. All the blue points are bounded by the  trade-off curves shown in (b).
		Holes in the plot indicate that the optimization algorithm did not find a local minimum consistent with the definition of a Pareto front. Panels (b), (c), and (d) present trade-off curves between pairs of observables in $\{\eta, v, \Delta \sigma\}$. The color dashed lines in plots with $\eta$ correspond to the minimum error and the black dashed lines correspond to the equilibrium error. The black line in panel (d) represents the scaling law expressed by Eq.~\eqref{eq:scalingvsigma}. Parameters are $\Delta E_r = 6.5$ and $\Delta E_w = 5.21$, which are the averages DNA binding energies for cognate and non-cognate pairings from ab-initio calculations  \cite{Sponer1996}, and $\Delta E_{\hyd}^{\tot} = 100$, which is sufficiently large value to recover results from the ideal Hopfield model. Results for $\Delta E_{\hyd}^{\tot} = 25$ and $\Delta E_{\hyd}^{\tot} = 12.5$ are qualitatively similar, see Appendix \ref{app1} (Figure~\ref{fig:Delta Ehyd = -12.5}).  
		\label{fig:hopfield_DEhyd=-100}}
\end{figure*}

\subsection*{Optimization algorithm}

We find the optimal solutions by a computational scheme based on multi-functional optimization \cite{miettinen2012nonlinear}. In particular, we systematically explore the parameter space using genetic algorithms \cite{deb2001multi} to identify feasible values of the observables that are compatible with the constraints and find the Pareto optimal front between $\eta$, $v$ and $\Delta \sigma$ as a function of the free parameters. We employ this method for different numbers of proofreading steps. 

In practice, the coefficients of the linear system satisfied by the concentrations $[m_i^x]$ can be of very different  magnitude, which might cause numerical instabilities. To counter this problem, we rescale Eqs.~\eqref{eq:hopfield_chemical_equations}, \eqref{eq:observables}, and \eqref{eq:nonlinear_constraints} by performing the change of variables
\begin{equation}
	\begin{aligned}
		K_\rightarrow^1 = & \frac{k^1}{k^{0}} \\
		K_\rightarrow^i = & \frac{k^i}{e^{ \sum_{j=1}^{i-1} \Delta E_{\conf}^j} k^{i-1,p}}  \quad \mbox{for } i = 2\ldots N\\
		K_\leftarrow^i = & \frac{e^{\Delta E_{\conf}^i} k^i}{e^{ \sum_{j=1}^{i} \Delta E_{\conf}^j}       k^{i,p}}  \quad \mbox{for } i = 1\ldots N-1\\
		[M_1^x] = & [m_1^x]  \\
		[M_i^x] = & \frac{[m_i^x] } {e^{ -\sum_{j=1}^{i-1} \Delta E_{\conf}^j}} \quad \mbox{for } i = 2\ldots N.
	\end{aligned} \label{eq:rescaling}
\end{equation}
With this prescription, the forward and the reverse rate constants are rescaled by the corresponding reverse proofreading rate constants, but without including the contributions from the binding energy and the hydrolysis energy. The concentrations are scaled by their corresponding Boltzmann factor, i.e., with their equilibrium value. We remark that the rescaling defined in Eq.~\eqref{eq:rescaling} is invertible, and the rescaled equations have therefore the same number of free parameters as the original equations.\\ 
More details about the rescaled form of the chemical equations are presented in the Appendix \ref{app1}.

\section{Results}
\label{sec5}
\subsection*{Pareto fronts}
We compute the Pareto optimal fronts for the generalized Hopfield model with different numbers of intermediate steps, see Figure~\ref{fig:hopfield_DEhyd=-100}(a). Theory \cite{Hopfield4135, NINIO1975587}  predicts that the error ranges from the minimum error
\begin{equation}\label{eq:minerr}
\eta_H=\frac{1}{1 + e^{N(\Delta E_r - \Delta E_w)}}
\end{equation}
up to the equilibrium error
\begin{equation}\label{eq:eqerr}
\eta_{eq}=\frac{1}{1 + e^{\Delta E_r - \Delta E_w}}.
\end{equation}
In all cases, the Pareto front between pairs of variables is a monotonic curve by definition, see Figure~\ref{fig:hopfield_DEhyd=-100}(b), (c), and (d). Speed-error and dissipation-error trade-off curves are defined in the error range $\eta \in[\eta_{eq},\eta_H]$ as expected.
In particular, at the equilibrium error the speed attains its maximum, and the dissipation tends to zero, see Figures~\ref{fig:hopfield_DEhyd=-100}(b), (c). The Pareto fronts for different $N$ appear to tend to a common limit $v\approx 1$ for $\eta\rightarrow\eta_{eq}$. In this limit, the reaction speed is limited by the rate of monomer binding $k^0$, while all other reaction steps are much faster. We note that the estimated dissipation per step steeply drops to zero for $\eta\rightarrow\eta_{eq}$, see Figure~~\ref{fig:hopfield_DEhyd=-100}(c). We expect the exact dependence of $\Delta \sigma$ on $\eta$ to be smooth around $\eta_{eq}$, as in exactly solvable models (see, e.g., \cite{Bennett1979,sartori2013kinetic}. We therefore expect the abrupt jump in Figure~~\ref{fig:hopfield_DEhyd=-100}(c) to be due to our finite numerical resolution and our choice of representing results in log scale.

The Pareto fronts between $\eta$ and $v$ and between $\eta$ and $\Delta \sigma$ substantially improve with the number of proofreading steps, see Figure~\ref{fig:hopfield_DEhyd=-100}(b) and ~\ref{fig:hopfield_DEhyd=-100}(c), respectively. Instead, the optimal front between $v$ and $\Delta \sigma$ barely depends on the number of proofreading steps, see Figure \ref{fig:hopfield_DEhyd=-100}(d). To understand this observation, we start from the full three-dimensional Pareto fronts represented in   Figure~\ref{fig:Delta Ehyd = -12.5}(a). If we now minimize $\Delta \sigma$ at fixed $v$ along these fronts, we find that the minimum is attained for large values of $\eta$ (i.e., close to $\eta_{eq}$). In this limit, proofreading is effectively inactivated, as its action would slow down the process and cause additional dissipation. The monomer incorporation pathway is then reduced to a linear one, for which varying the number of steps at fixed chemical driving has little effect on the speed-dissipation Pareto front.

\begin{figure}
	\centering
	\includegraphics[width=\linewidth]{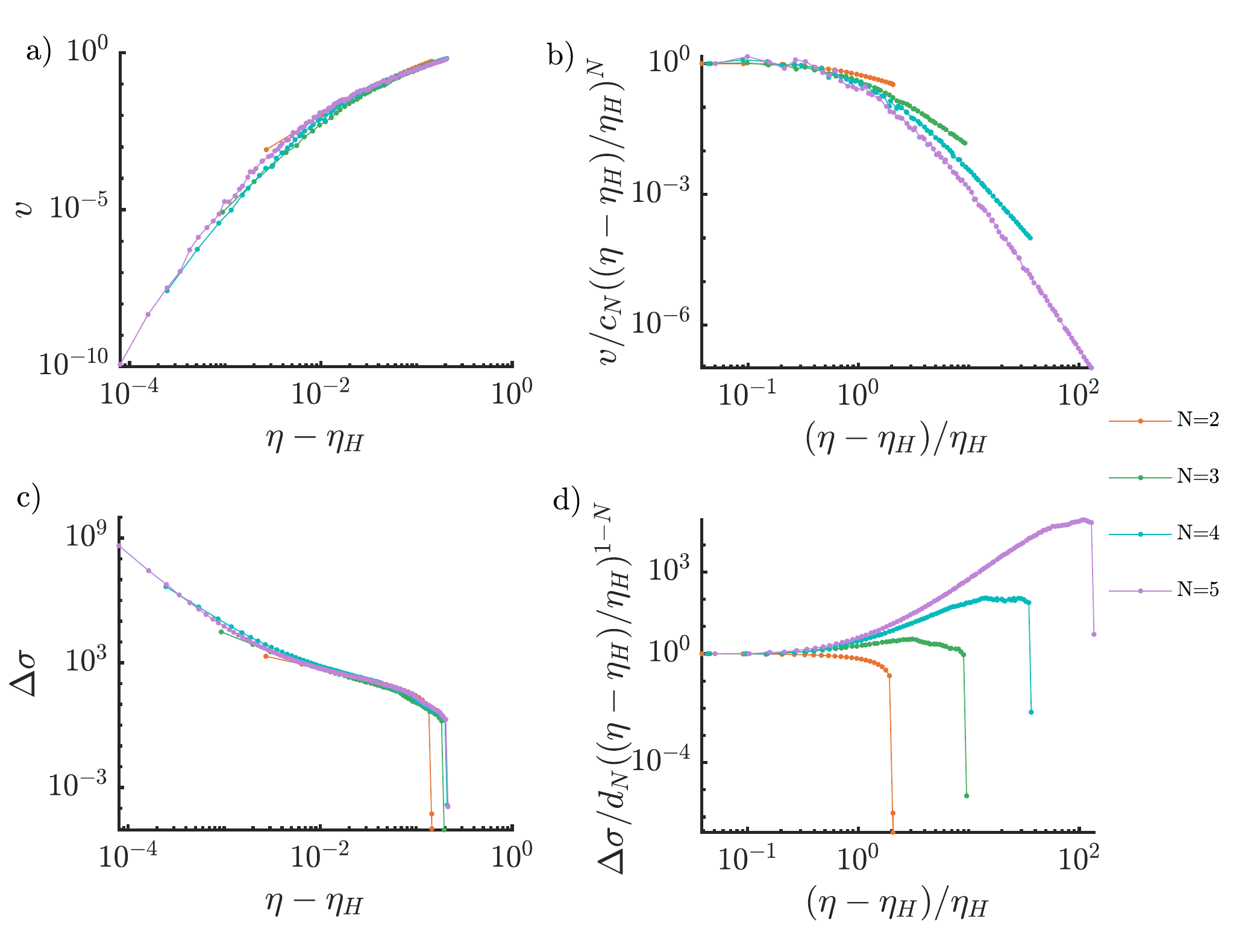}
	\caption{Scaling relation between error speed, and error dissipation in two-dimensional Pareto fronts for different $N$. Plotting as a function of $(\eta-\eta_{H})$, we obtain similar curves for the Pareto fronts between (a) $\eta$ and $v$ and (c) $\eta$ and $\Delta\sigma$. Panels (b) and (d) show data collapses according to Eqs.\eqref{eq:collapse1} and \eqref{eq:collapse2}, respectively.  In both cases, we find a data collapses for small error rates. In all curves, we fix $\Delta E_{\hyd}=100$.}
	\label{fig:scaling_relations}
\end{figure} 

\subsection*{Scaling Laws}

We study the behavior of the speed and dissipation on the Pareto front for errors close to the minimum error $\eta_H$ given by Eq.~\eqref{eq:minerr}. Plotting velocity and dissipation as a function of $(\eta-\eta_H)$ leads to a good data collapse, see Figure~\ref{fig:scaling_relations}\sm{(b, d)}. However, a scrutiny of the data collapse in (a) and (c) shows that the slope is different for different $N$. In fact, we find that, for small errors, the behaviors of the velocity and the dissipation are well described by scaling relations
\begin{equation}\label{eq:collapse1}
	v\sim \left( \frac{\eta-\eta_{H}}{\eta_{H}} \right)^N
\end{equation}
and 
\begin{equation}\label{eq:collapse2}
	\Delta \sigma \sim \left( \frac{\eta-\eta_{H}}{\eta_{H}} \right)^{1-N},
\end{equation}
see Figure~\ref{fig:scaling_relations}(b) and ~\ref{fig:scaling_relations}(d), respectively.

\begin{figure}
	\centering
	\includegraphics[width= \linewidth]{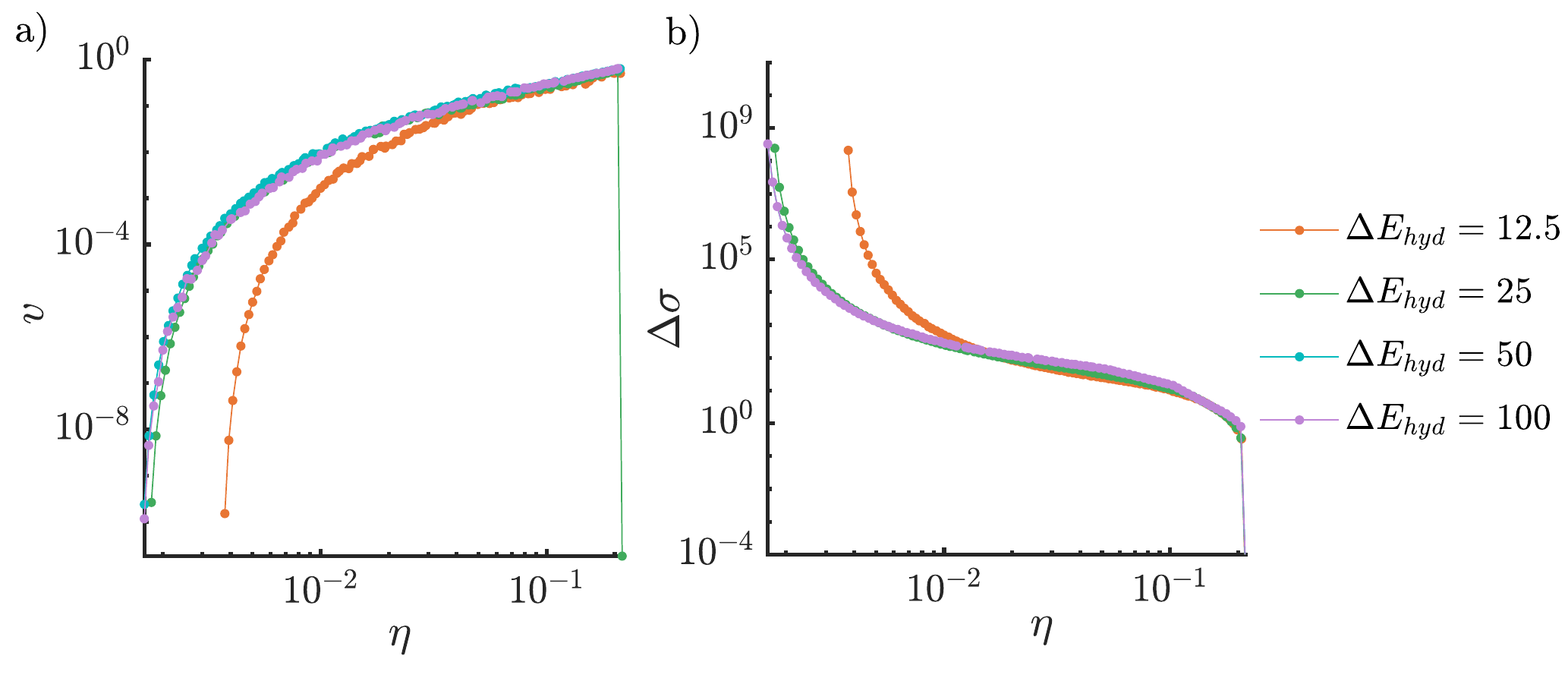}
	\caption{Higher hydrolysis energy $\Delta E_{\hyd}$ yields better Pareto fronts. (a), (b)  Trade-offs between pairs of observables in $\{\eta, v, \Delta \sigma\}$ for different $\Delta E_{\hyd}$. In all cases, the Pareto fronts improve at higher hydrolysis energy up to a certain limit when the $\Delta E_{\hyd}$ (e.g  $\Delta E_{\hyd}=12.5$)  is not sufficient for the enzyme to achieve the high energy state. All curves in this figure are for $N=5$. }
	\label{fig:scaling_energy}
\end{figure}  

Our data suggest a scaling law between $v$ and $\Delta \sigma$ of the form 
\begin{equation}\label{eq:scalingvsigma}
	v \sim (\Delta \sigma)^\gamma
\end{equation}
with $\gamma = 1.85 \pm 0.05$, at least for a range of small values of $\Delta \sigma$, see Figure~\ref{fig:hopfield_DEhyd=-100}.

\begin{figure}
	\centering
	\includegraphics[width = \linewidth]{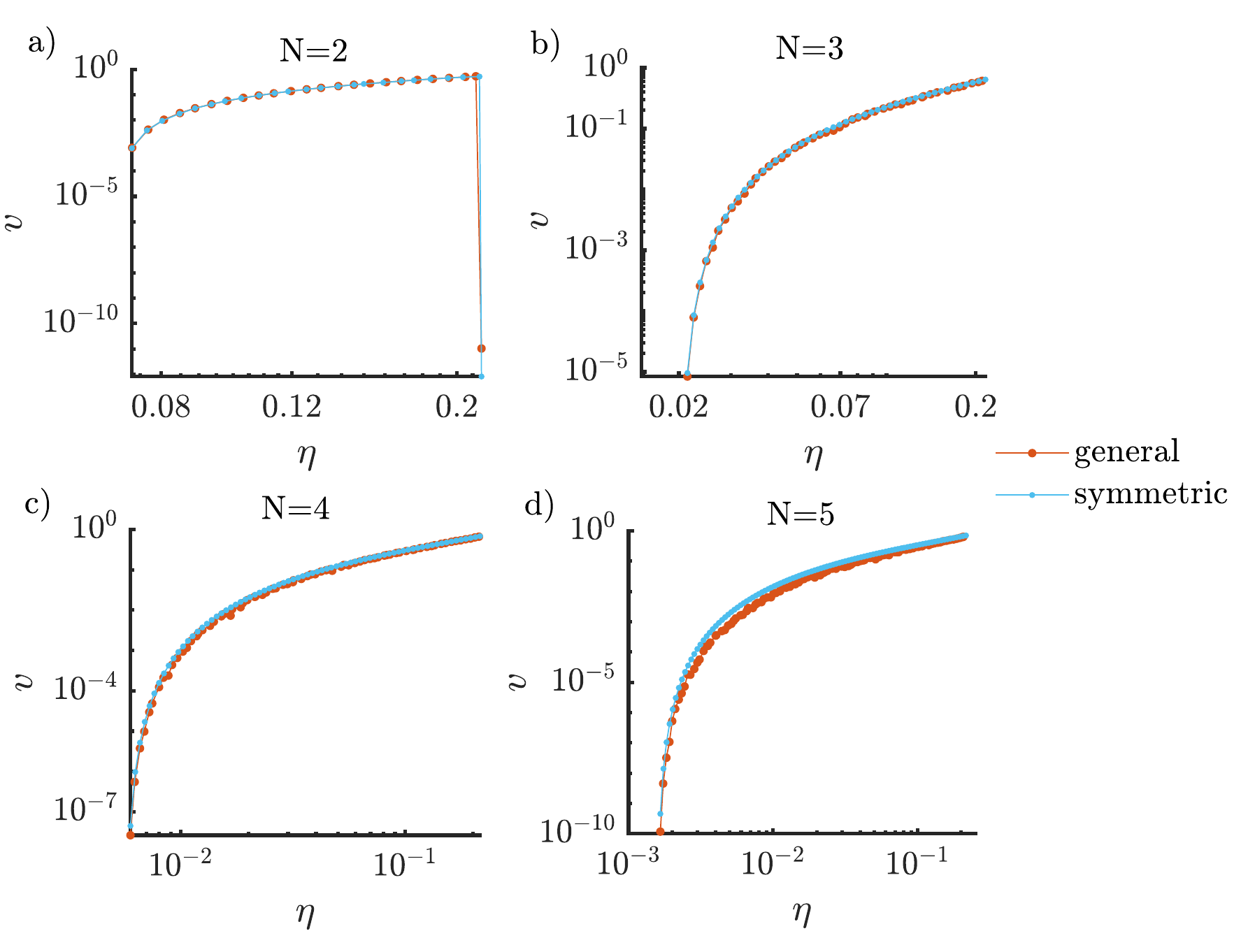}
	\caption{Comparison of error-speed trade-off curves for symmetric and general Hopfield models.} 
	\label{fig:comparison_general_and_symmetric_equations}
\end{figure}

Both speed-error and dissipation-error Pareto fronts improve with $\Delta E_{\hyd}$ and seems to saturate for large $\Delta E_{\hyd}$, see Figure~\ref{fig:scaling_energy}. In particular, an increase from $\Delta E_{\hyd}=12.5$  to $\Delta E_{\hyd}=25$ leads to significant improvements. More extensive numerical results on the Pareto fronts for $\Delta E_{\hyd}=12.5$ are presented in Appendix \ref{app_energy}. On the other hand, a further increase from $\Delta E_{\hyd}=25$ to  $\Delta E_{\hyd}=100$ does not lead to appreciable changes. This result shows that $\Delta E_{\hyd}=25$ (a value on the order of the free energy of ATP hydrolysis) can be already considered as very large, at least for the values of $N$ that we considered.

\subsection*{Symmetric choice of parameters.}

The rescaling of the parameters introduced in Eq.~\eqref{eq:rescaling} serves to have rates of comparable order of magnitude. In this section, we explore what happens if we assume that the rescaled optimal parameters are independent of the reaction step.  In particular, we assume that  $K^i_{\to}=K_{\to}$ and $K^i_{\leftarrow}=K_{\leftarrow}$. We also assume that the system consume equal amount of hydrolysis energy  at each step, $\Delta E^i_{\hyd}=\Delta E_{\hyd}$ for $i=1,\dots N$. As a result, we are left with only two free parameters, $K_{\to}$ and $K_{\leftarrow}$, irrespective of $N$.  We find that, the Pareto fronts obtained under this approximation are similar to the general one, see  Figure~\ref{fig:comparison_general_and_symmetric_equations}. In particular, for $N=2$, the Pareto fronts between speed and error in the symmetric and general model are indistinguishable in our numerical simulations. The difference between the Pareto fronts in the two models increases with the number of steps.

\section{Conclusions}

In this work, we numerically studied the Pareto optimal fronts between speed, error rate, and dissipation in biological error correction. Our algorithm, based on multi-object optimization theory, permits to reconstruct the Pareto front with high accuracy. 

Our results demonstrate aspects of error corrections that would be difficult to reveal with other theoretical methods. For example, we find that the error/speed and error/dissipation trade-offs strongly depend on the number of proofreading steps. In contrast, the trade-off between speed and dissipation is rather insensitive to the number of steps. In certain regimes, we found that speed, error rate, and dissipation along the Pareto front are characterized by non-trivial scaling laws. Finally, we have found that a rescaling of the chemical equations, that we introduce to obtain more stable numerical results, suggests a symmetry assumption on the optimal rates that leads to a good approximation of the Pareto front. This result is a potentially useful hint for future theoretical approaches. 

We focused our study on a generalization of the Hopfield model, where the discrimination in the main incorporation pathway is only present in backward rates. In many real error correction networks, such as those implemented by ribosomes, this discrimination strategy is complemented by forward discrimination, i.e., discrimination based on energy barrier differences \cite{doi:10.1098/rstb.2016.0182,EHRENBERG1980333,banerjee2017elucidating}. The presence of forward discrimination leads to more complex scenarios, where the choice of parameters to be optimized and global constraints might play a delicate role. Understanding these cases using the numerical approach outlined in our work is an interesting venue of investigation for future studies.

\appendix

\section{Computational details of error-speed-dissipation trade-off}
\label{app1}

The rescaled version of Eq.~\eqref{eq:hopfield_chemical_equations} at steady state are expressed by
\begin{eqnarray}
	\begin{aligned}
			0 = & 1+ \left( -e^{-\Delta E_x} -e^{\Delta E_{\hyd}^1}K_\rightarrow^{1} \right) M_1^{x} + K_\rightarrow^{1} M_2^{x}\\
			0 = & 1+  e^{\Delta E_{\hyd}^{i-1}} K_\leftarrow^{i-1} [M_{i-1}^x] \\
			&  + \left(-e^{-\Delta E_x}- K_\leftarrow^{i-1}
			- e^{\Delta E_{\hyd}^{i}}K_\rightarrow^{i}\right)[M_i^x]\\ 
			& + K_\rightarrow^{i} [M_{i+1}^x] \\
			& \mbox{for } i = 2\ldots N-1\\
			0 = & 1+ e^{\Delta E_{\hyd}^{N-1}}K_\leftarrow^{N-1} [M_{N-1}^x] \\
			&  +\left(- e^{-\Delta E_x} -K_\leftarrow^{N-1} - K_\rightarrow^{N} \right)[M_N^x]\\
	\end{aligned}\label{eq:_rescaled_hopfield_chemical_equations}
\end{eqnarray}
Eqs.~\eqref{eq:_rescaled_hopfield_chemical_equations} are linear in the rescaled variables $[M_i^x]$. To compute the steady concentrations, we express Eq.~\eqref{eq:_rescaled_hopfield_chemical_equations} in matrix form
\begin{equation}
	\mathbf{P}\cdot \mathbf{M}= \mathbf{Q}\label{eq:pmatrix}
\end{equation}
where $\mathbf{M}=\{ [M_1^x]...[M_N^x]\}$, $\mathbf{Q}$ is a constant vector with all entries equal to $-1$, and $\mathbf{P}$ is the matrix of coefficients of the linear system. For example, for $N=2$, the matrix $\mathbf{P}$ reads 
\begin{equation}
\mathbf{P} = \begin{pmatrix}
	- e^{-\Delta E_x}- e^{\Delta E_{\hyd}^{1}}K_\rightarrow^{1} &     K_\rightarrow^{1} \\
	e^{\Delta E_{\hyd}^{1}}K_\leftarrow^{1} &    - e^{-\Delta E_x}- K_\leftarrow^{1} - K_\rightarrow^{2} .
\end{pmatrix}
\end{equation}

Next, we invert the rescaled relations to express the original parameter as a function of the rescaled parameters $K_{\rightarrow}^i$ and $K_{\leftarrow}^i$. From Eq.~\eqref{eq:rescaling} we obtain the inverse relations
\begin{equation}
    k_{\rightarrow}^{i,p}=k_{\rightarrow}^{0}\prod_{j=1}^i \frac{K_{\rightarrow}^{j}}{K_{\leftarrow}^{j}}
\end{equation}

\begin{equation}
    k^1=k^0 K_\rightarrow ^1
\end{equation}

\begin{equation}
	\begin{aligned}
   k_{\rightarrow}^{i}= &\, e^{\sum_{j=1}^{i-1}\Delta E_{\conf}^{j}} K_{\leftarrow}^{i} k_{\rightarrow}^{0}\prod_{j=1}^i \frac{K_{\rightarrow}^{j}}{K_{\leftarrow}^{j}}\\
   & \mbox{for } i = 2\ldots N-1\\
   k^N= &\, k^0 e^{\sum_{j=1}^{i-1}\Delta E_{\conf}^{j}}\frac{  \prod_{j=1}^N K_{\rightarrow}^{j}}{  \prod_{j=1}^{N-1}K_{\leftarrow}^{j}}
   \end{aligned}
\end{equation}
	

We can now write the observables $\eta$, $v$, and $\Delta \sigma$ in terms of the rescaled rates and concentrations:
\begin{equation}
	\begin{aligned}
		\eta = &\, \frac{[M_N^w]}{[M_N^w] + [M_N^r]}\\
		v  = &\, k^0\frac{  \prod_{j=1}^N K_{\rightarrow}^{j}}{  \prod_{j=1}^{N-1}K_{\leftarrow}^{j}}\left([M_N^r] + [M_N^w] \right)\\
		\Delta \sigma =&\, \frac{\sigma^r+\sigma^w}{v}
	\end{aligned}\label{eq:rescaled_observables}
\end{equation}
where 


\begin{equation}
	\begin{aligned}
 \sigma_n^{x}=& k^0\left( \sum_{i = 1}^{N-1} e^{\Delta E_{\hyd}^{i}}K_\rightarrow^{i} \log\left[ \frac{e^{-\Delta E_{\hyd}^{i}} M_i^x}{M_{i+1}^x}\right]  \right)\\ \nonumber
		& \times \left( \left(M_i^x -e^{\Delta E_{\hyd}^{i}} M_{i+1}^x\right) \prod_{j = 1}^{i-1}\frac{K_\rightarrow^{j}}{K_\leftarrow^{j}} \right)\\ \nonumber
		&+k^0 \left( \sum_{i = 0}^{N-1}-\log[\frac{e^{-\Delta E_x}}{M_{i+1}^x}] (-1 + e^{-\Delta E_x} M_{i+1}^x \right)\\ \nonumber
		& \times \prod_{j = 1}^i\frac{K_\rightarrow^{j}}{K_\leftarrow^{j}}
	\end{aligned}
\end{equation}

Finally, we feed Eq.~\eqref{eq:pmatrix} and Eq.~\eqref{eq:rescaled_observables} to the MATLAB function \texttt{gamultiobj} to obtain  three-dimensional Pareto fronts for $\eta$, $v$, and $\Delta\sigma$.  The function \texttt{gamultiobj} uses genetic algorithms \cite{deb2001multi} and numerical evaluations of Eq.~\eqref{eq:rescaled_observables} to find the combination of free parameters corresponding to the dominant configurations of $\eta$, $v$, and $\Delta\sigma$.

The rescaled problem can still generate badly scaled matrices in some rare instances, which we further mitigate by implementing arbitrary precision computations with the Advanpix multi-precision package for MATLAB \cite{mct2015}. In this way, we efficiently obtain numerically stable results. 

\section{Results for different values of the hydrolysis energy. }
\label{app_energy}

\begin{figure}
\centering
\includegraphics[width=\linewidth]{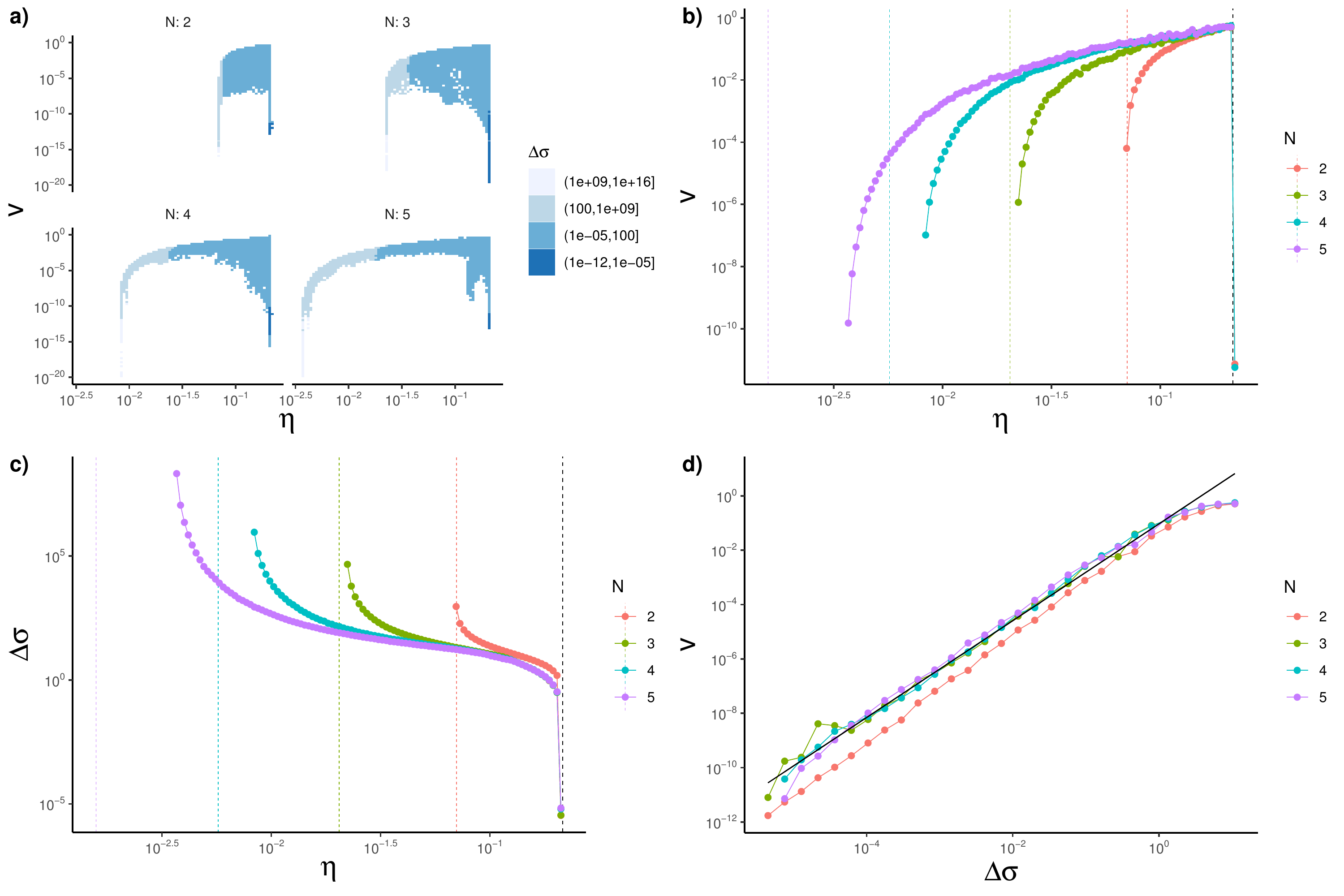}
\includegraphics[width=\linewidth]{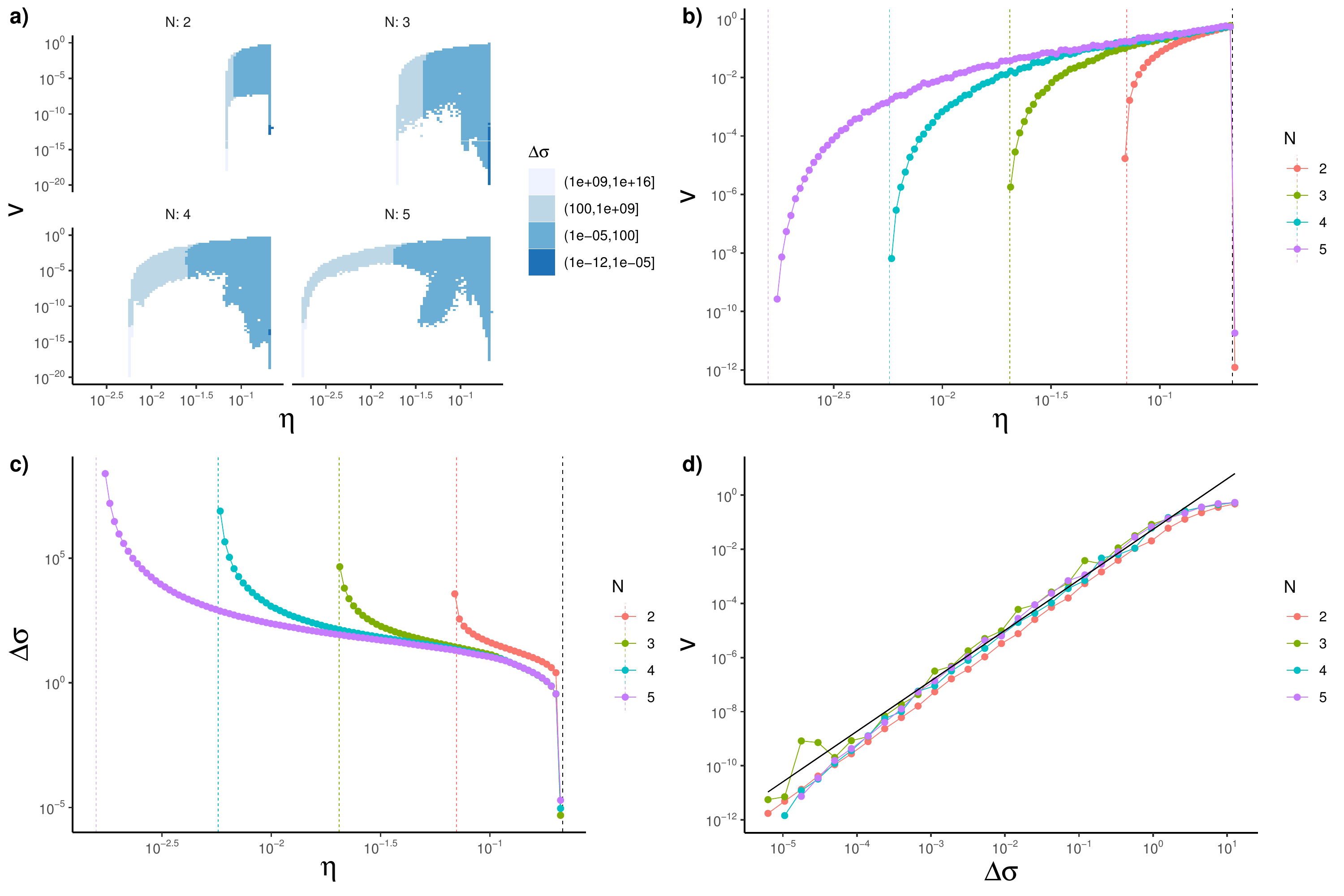}
\caption{Pareto optimal fronts (a-d) for $\Delta E_{\hyd} = 12.5$ and (e-h) for $\Delta E_{\hyd} = 25$ (a),(e) 3-Dimensional Pareto optimal front between $\eta$, $v$ and $\Delta \sigma$ for different number of steps  $N$. (b), (c), (d), (f), (g), and (h) marginalized trade-off curves between pairs observables in $\{\eta, v, \Delta \sigma\}$. Dashed lines in plots with $\eta$ correspond to the ideal error of the Hopfield model with $N$ intermediate steps.}
\label{fig:Delta Ehyd = -12.5}
\end{figure}

We computed the Pareto front by setting the hydrolysis energy $\Delta E_{\hyd}^{\tot}=12.5$ and $\Delta E_{\hyd}^{\tot}=25$ (see Figure~\ref{fig:Delta Ehyd = -12.5}) Figure~\ref{fig:Delta Ehyd = -12.5} is qualitative similar to the case for $\Delta E_{\hyd}^{\tot}=100$  (see Figure~\ref{fig:hopfield_DEhyd=-100}) and agree with our observation that additional proofreading steps improve Pareto fronts. However, for $\Delta E_{\hyd}^{\tot}=12.5$ Figure~\ref{fig:Delta Ehyd = -12.5}(a-d), minimum errors obtained from the simulation are substantially larger than the minimum errors defined in Eq.~\ref{eq:minerr}, in particular for $N>3$.

\begin{acknowledgments}
We thank Deepak Bhat for useful discussions. SP was supported by JSPS KAKENHI Grant Number JP18K03473 and by the Okawa Foundation (Grant Number 21-01).
\end{acknowledgments}
\bibliography{references}

\end{document}